\documentclass[aps,prl,twocolumn,superscriptaddress]{revtex4-1}
\usepackage{lipsum}

\usepackage{graphicx}
\usepackage{ulem}
\usepackage{color}

\usepackage{xspace}

\usepackage{amsmath}
\usepackage{siunitx}
\usepackage{miller}

\usepackage[version=3]{mhchem}

\newcommand{\red}{}

\newcommand{\reza}{\AA^{-1}}

\newcommand{\tn}{$T_\mathrm{N}$\xspace}
\newcommand{\treo}{$T_{\mathrm{reo}}$\xspace}

\newcommand{\tc}{$T_{\mathrm{c}}$\xspace}

\newcommand{\bfca}{Ba(Fe$_\text{1-x}$Co$_\text{x}$)$_\text{2}$As$_\text{2}$\xspace}
\newcommand{\bfcaop}{Ba(Fe$_\text{0.94}$Co$_\text{0.06}$)$_\text{2}$As$_\text{2}$\xspace}

\newcommand{\bfcaud}{Ba(Fe$_\text{0.955}$Co$_\text{0.045}$)$_\text{2}$As$_\text{2}$\xspace}

\newcommand{\nfca}{NaFe$_\text{1-x}$Co$_\text{x}$As\xspace}
\newcommand{\bkfa}{Ba$_\text{1-x}$K$_\text{x}$Fe$_\text{2}$As$_\text{2}$\xspace}

\newcommand{\bfa}{BaFe$_\text{2}$As$_\text{2}$\xspace}

\newcommand{\e}{\text{e}}

\newcommand{\reslib}{\textsc{Reslib}\xspace}

\newcommand{\bnfax}{Ba$_\text{1-x}$Na$_\text{x}$Fe$_\text{2}$As$_\text{2}$\xspace}

\newcommand{\fescan}{\ref{fig:escans}\xspace}

\newcommand{\ftdep}{\ref{fig:tdep}\xspace}

\newcommand{\bs}[1]{\boldsymbol{#1}}
\newcommand{\lb}{\left(\xspace}
\newcommand{\rb}{\right)\xspace}
\newcommand{\q}{$\bs{Q}$\xspace}

\newcommand{\repl}[2]{{#2}}

\usepackage{xifthen}
\newcommand{\headline}[1]{{\boldmath\textbf{#1}}}
\newcommand{\hlcaption}[2][]{
	\caption[{#1}]{
		\ifthenelse{\isempty{#1}}
		{#2}
		{\headline{#1}#2}
	}
}

\begin{document}

\advance\vsize by 2 cm

\title{Strong Spin Resonance Mode associated with suppression of soft magnetic ordering in Hole-doped Ba$_\text{1-x}$Na$_\text{x}$Fe$_\text{2}$As$_\text{2}$ }

\author{F. Wa\ss{}er}
\affiliation{II. Physikalisches Institut, Universit\"{a}t zu K\"{o}ln, Z\"{u}lpicher Stra\ss{}e 77, D-50937 K\"{o}ln, Germany}

\author{J.-T. Park}
\affiliation{Heinz Maier-Leibnitz Zentrum (MLZ), Technische Universit\"at M\"unchen, Lichtenbergstr. 1, D-85748 Garching, Germany}

\author{S. Aswartham}
\affiliation{%
Leibniz-Institut f\"ur Festk\"orper- und Werkstoffforschung Dresden, Helmholtzstra\ss{}e 20, D-01069 Dresden, Germany}

\author{S. Wurmehl}
\affiliation{%
Leibniz-Institut f\"ur Festk\"orper- und Werkstoffforschung Dresden, Helmholtzstra\ss{}e 20, D-01069 Dresden, Germany}
\affiliation{Institut f\"ur Festk\"orperphysik, Technische Universit\"at Dresden, D-01171 Dresden, Germany}

\author{Y. Sidis}
\affiliation{Laboratoire L\'eon Brillouin, C.E.A./C.N.R.S., F-91191 Gif-sur-Yvette CEDEX, France}

\author{P. Steffens}
\affiliation{%
Institut Laue Langevin, 71 avenue des Martyrs, 38000 Grenoble, France}

\author{K. Schmalzl}
\affiliation{%
J\"ulich Centre for Neutron Science, Forschungszentrum J\"ulich GmbH, Outstation at Institut Laue-Langevin, 71 avenue des Martyrs, 38000 Grenoble, France}


\author{B. B\"uchner}
\affiliation{%
Leibniz-Institut f\"ur Festk\"orper- und Werkstoffforschung Dresden, Helmholtzstra\ss{}e 20, D-01069 Dresden, Germany}
\affiliation{Institut f\"ur Festk\"orperphysik, Technische Universit\"at Dresden, D-01171 Dresden, Germany}

\author{M. Braden}
\email{braden@ph2.uni-koeln.de} \affiliation{II. Physikalisches Institut, Universit\"{a}t zu K\"{o}ln, Z\"{u}lpicher Stra\ss{}e 77,
D-50937 K\"{o}ln, Germany}

\date{\today}


\begin{abstract}

Spin-resonance modes (SRM) are taken as evidence for magnetically driven pairing in Fe-based superconductors, but their character remains poorly understood. The broadness, the splitting and the spin-space anisotropies of SRMs contrast with the mostly accepted interpretation as spin excitons. We study hole-doped Ba$_{1-x}$Na$_x$Fe$_2$As$_2$ that displays a spin reorientation transition. This reorientation has little impact on the overall appearance of the resonance excitations with a high-energy isotropic and a low-energy anisotropic mode. However, the strength of the anisotropic low-energy mode sharply peaks at the highest doping that still exhibits magnetic ordering resulting in the strongest SRM observed in any Fe-based superconductor so far. This remarkably strong SRM is accompanied by a loss of about half of the magnetic Bragg intensity upon entering the SC phase. Anisotropic SRMs thus can allow the system to compensate for the loss of exchange energy arising from the reduced antiferromagnetic correlations {\red within the SC state.}


\end{abstract}

\maketitle

\section*{Introduction}

Superconductivity (SC) in iron-based superconductors emerges upon suppressing an orthorhombic and antiferromagnetic (o-AFM) state \cite{Kamihara2008,Dai2015a,Hirschfeld2011} and
the highest \tc values are realized, when AFM order is completely suppressed and when spin fluctuations are expected to be the strongest.
These observations inspired theories with spin-fluctuation-mediated pairing and a SC order parameter that changes sign between the hole and electron pockets of the {\red Fermi surface} \cite{Hirschfeld2011}.
A hallmark prediction of such theory is a spin exciton appearing below \tc at $E_{res} < 2\Delta_{SC}$, with $\Delta_{SC}$ the SC gap energy \cite{Korshunov2008,Korshunov2016,Hirschfeld2011}.
Experimentally, numerous inelastic neutron scattering (INS) experiments observe spin-resonance modes (SRM) emerging in the SC state at energies below the electron-hole continuum \cite{Christianson2008,Inosov2009,Zhang2011,Lee2013,Qureshi2012,Wang2016f}.

In a simple picture the residual magnetic correlations are likely to generate a collective S=1 mode in the SC state, which should be single and isotropic in spin-space, i.e. the magnetic polarization should be identical along all directions. However, polarised INS experiments in \bfcaop (optimally electron doped) reveal a more complex
character \cite{Steffens2013} by separating the distinct channels in spin space. In addition to a broad SRM that is isotropic and lies at a higher energy corresponding
to $E_{\mathrm{res}}\sim  5.3 k_BT_{\mathrm{c}}$ there is a sharp low-energy SRM-1 that is anisotropic in spin space. Qualitatively the same superposition of a low-energy anisotropic
feature SRM-1 and a broader isotropic one at higher energies, SRM-2, has been established in various FeAs-based superconductors: in \bfcaop (optimally electron doped) \cite{Steffens2013,Wasser2017}, in \nfca \cite{Zhang2014,Song2017}, in P-doped BaFe$_2$As$_2$ \cite{Hu2017}, in Ni-doped BaFe$_2$As$_2$ \cite{Luo2013}, and in hole-doped BaFe$_2$As$_2$ \cite{Zhang2013b,Qureshi2014}.
Various attempts were made to attribute the split and anisotropic SRMs
to either persisting antiferromagnetic correlations and their corresponding excitations \cite{Knolle2011,Lv2014,Wang2016c} or to an orbital-selective pairing \cite{Yu2014,Korshunov2016}. However,
none of these scenarios can account for the existence and the character of the split resonance modes in a large variety of materials extending into the overdoped
range \cite{Wasser2017,QureshiBraden2012,Wang2013}. The spin-space anisotropy must be attributed to finite spin-orbit coupling, whose relevance in iron-based superconductors
is documented in the large anisotropy gaps of the magnon dispersion in the o-AFM phase \cite{Qureshi2012,Song2013}, in the splitting of otherwise degenerate electron energy bands \cite{Borisenko2016} and in the spin reorientation transition in hole-underdoped \bfa that restores tetragonal symmetry \cite{Avci2014,Wasser2015,Boehmer2015,Taddei2016a,Scherer2018}.

\begin{figure}[t]
\begin{center}
	\includegraphics*[width=1.1\columnwidth]{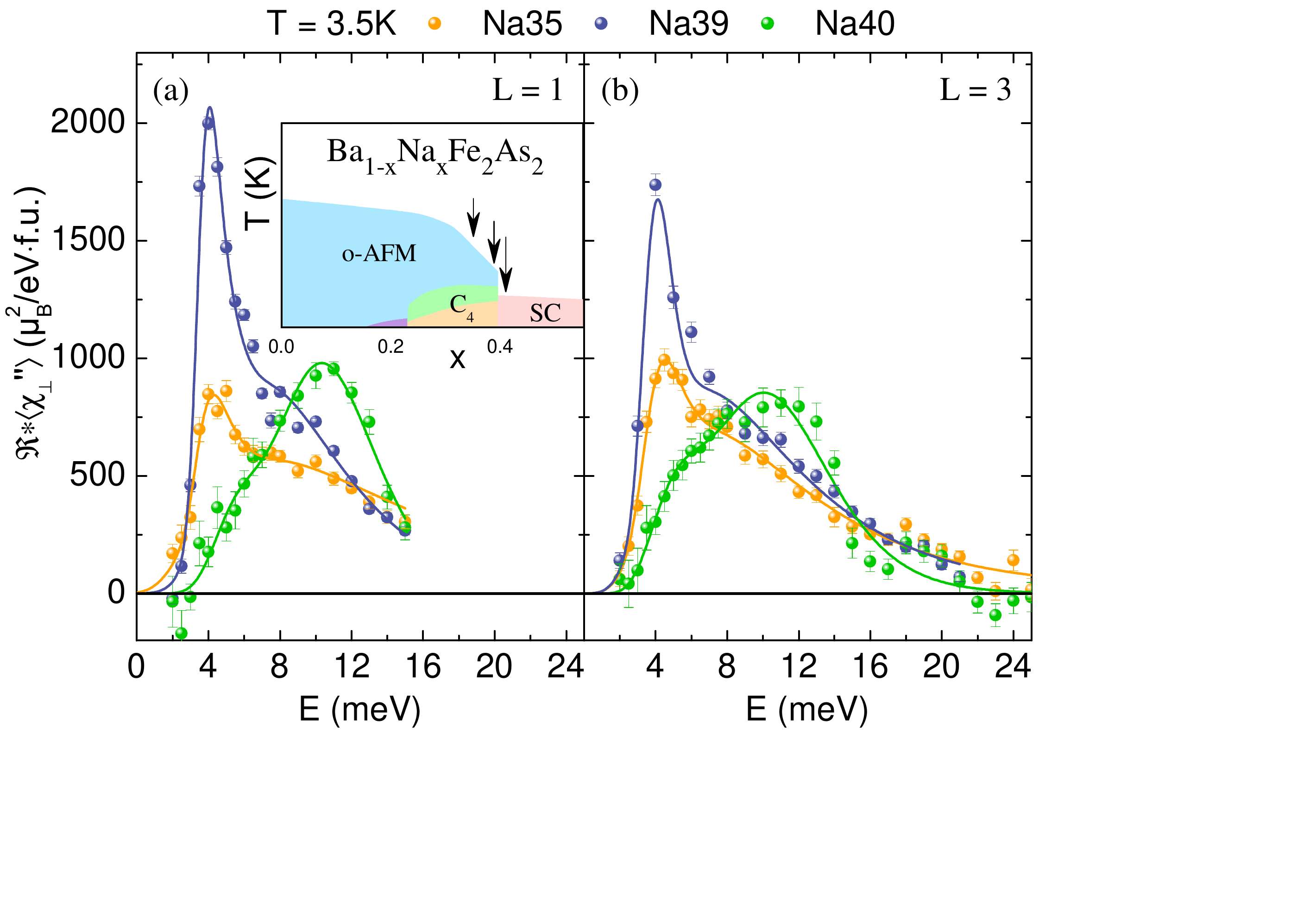} 
\end{center}
\vskip-1.5cm	
\hlcaption[Spin resonance modes in the superconducting phases.]
	{ INS spectra obtained at the AFM Bragg positions (0.5,0.5,L) with L=1,3 for the three samples studied. Data were corrected for background, for the Bose and form factors and normalized to absolute units, see methods section. The spectra were fitted with two log-normal functions corresponding to the two SRM's at 4 and 8-10meV.
The inset indicates {\red the location in the phase diagram of \bnfax \ for the three samples studied}.			 }
	\label{fig:res-sc}
\end{figure}

Concerning the split SRM's, the two best studied series are both electron doped: \bfca and \nfca  \cite{Steffens2013,Wasser2017,Song2017,Wang2017,Zhang2014}.
For underdoped \bfca (x = \SI{4.5}{\%}), nematic and AFM order with sizeable moments precede SC resulting in a large gap in the longitudinal spin fluctuations (polarized along
the ordered moment in orthorhombic $a$ direction). In consequence only anisotropic resonance modes emerge in the {\red transverse} channels \cite{Wasser2017},
which clearly disagree with the isotropic spin-exciton model \cite{Hirschfeld2011}.
At optimal Co doping (x = \SI{6}{\%}), an anisotropic low-energy mode is found in addition to an isotropic one at higher energies \cite{Steffens2013,Wasser2017}, while for Ni overdoping (x = \SI{10.9}{\%}) only the isotropic \repl{one}{part} persists \cite{Wang2016c}.
In \nfca the SRM evolves in a similar way with increasing doping; the intense and anisotropic low-energy part is gradually suppressed following the suppression of the AFM ordering, while the part at higher energies is isotropic throughout the entire SC dome \cite{Zhang2016a}.

For hole doped \bkfa \, INS studies were focusing on samples with optimal and overdoped composition \cite{Zhang2011,Qureshi2014,Lee2016a,Horigane2016,Wang2013a}, where AFM order is completely suppressed \cite{Boehmer2015}, or on higher energies \cite{Murai2018}. For hole-doped Sr$_{1-x}$Na$_x$Fe$_2$As$_2$ only a single very underdoped composition with low {\red $T_c$} was studied \cite{Guo2019}, which revealed only very weak signature of SRM's.
In the underdoped hole-doped regime, there is a second magnetic phase appearing in the o-AFM dome  at \treo $<$ \tn , where the magnetic moments reorient from in-plane towards out-of-plane alignment  \cite{Wasser2015,Allred2015,Taddei2016a}, c.f. schematic phase diagram in Fig.~1. For this spin-reoriented phase
a magnetic double-$\bs{k}$ structure has been reported leaving half of the Fe sites non-magnetic \cite{Allred2016,Mallett2015a} and tetragonal ($C_4$) symmetry is recovered \cite{Boehmer2015,Wang2016}. Following common use in the literature  we label this phase as C$_4$.
Recent studies of the local structure suggest that these C$_4$ phases with an average tetragonal structure exhibit very strong orthorhombic
fluctuations on short length and time scales \cite{Frandsen2017}.
In order to access this region where magnetism is particularly soft we explored the spin excitations in \bnfax by polarised and unpolarised INS for two slightly underdoped samples x = 0.35 (Na35, \tc=26\,K) and x = 0.39 (Na39, \tc=29\,K) and one optimally doped sample with x = 0.4 (Na40, \tc=34\,K), see methods section.
While the two former compounds exhibit magnetic ordering (\tn = 70 and 61\,K)  and the spin reorientation (\treo = 46 and 44.5\,K), Na40 does not show any evidence of magnetic ordering. In spite of the minor doping differences these samples exhibit fully different low-energy SRM's while the high-energy SRM
varies continuously with doping and \tc . This clearly indicates the close connection of the sharp low-energy SRM with persisting magnetic ordering.

\begin{figure}[t]
\vskip-3.5cm
	\includegraphics*[width=1.2\columnwidth]{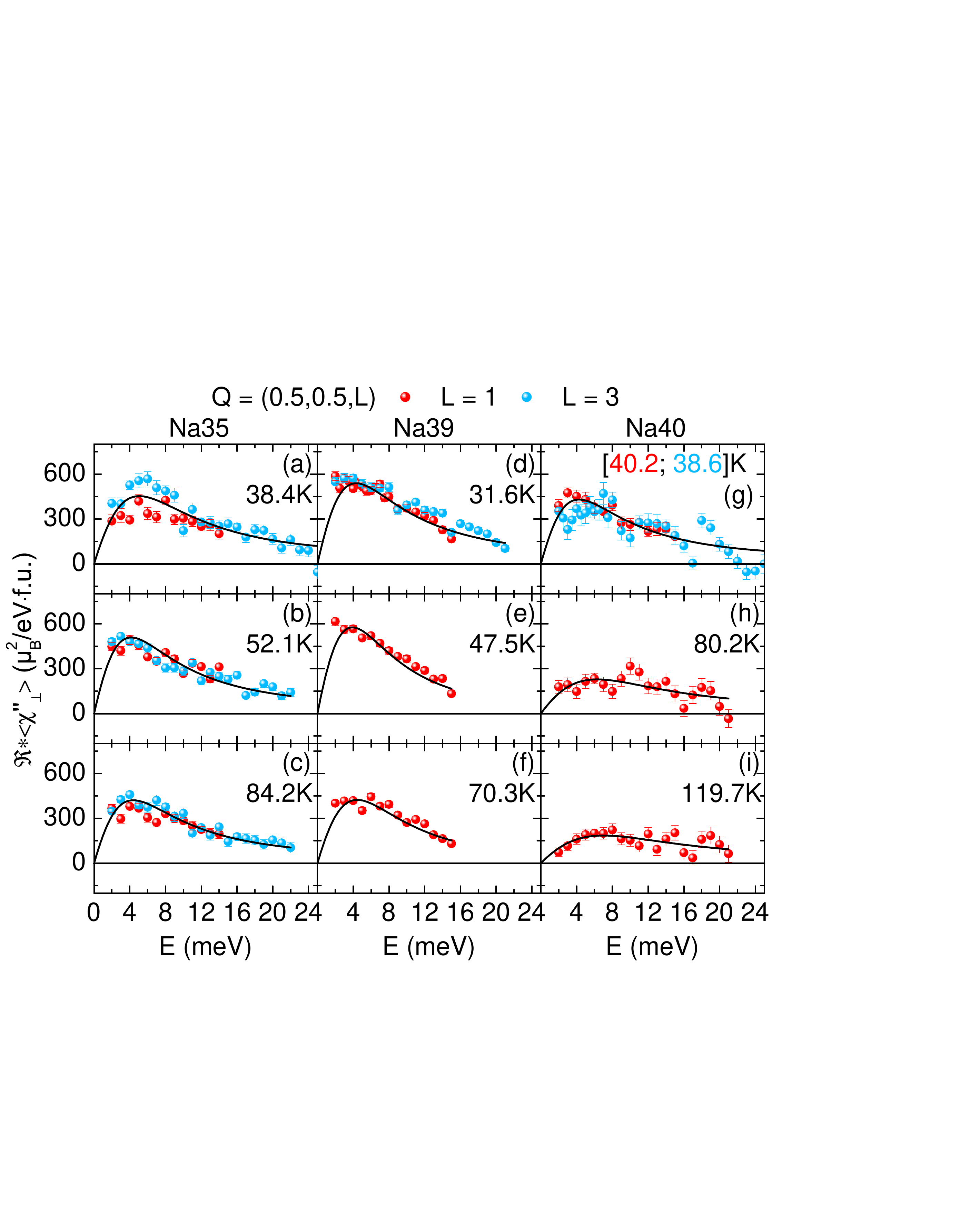} 
\vskip-2.5cm
	\hlcaption[Magnetic excitations at the AFM Bragg positions in the normal state]
	{ \textbf{(a)-(i)} Magnetic scattering presented in absolute units for the three samples recorded at (0.5,0.5,1) and (0.5,0.5,3) at various temperatures.
The NS response is described by a single relaxor function convoluted with the instrumental resolution.
		}
	\label{fig:escans}
\end{figure}

\section*{Results}

Fig. 1 shows our main result comparing the magnetic response in the SC state for the three concentrations Na35, Na39 and Na40, while Fig. 2 compares the magnetic response of these
materials at various temperatures in the normal state (NS).
Magnetic INS averages the two components of the generalized dynamic magnetic susceptibility, ${\chi}''\lb \bs{Q},E \rb$, perpendicular to the scattering vector, $\bs{Q}$.
The INS data were corrected for the background, the Bose statistics and higher-order contaminations and transformed into absolute units as described in the methods section. Since the susceptibility can be anisotropic, the signal corresponds to the average of the two perpendicular components {\red convoluted} with the resolution function of the instrument,
$\Re \star \langle {\chi''_{\perp}}\rangle  \lb \bs{Q},E \rb$.
In agreement with previous INS studies on various
electron and hole doped compounds \cite{Wasser2017}, the magnetic response in the SC state can be described by a SRM-2 at higher energy, $\sim$8-10\,meV, and a sharper SRM-1 at lower energy, $\sim$4\,meV.
While the higher SRM-2 varies hardly with the doping, the low-energy SRM-1 exhibits a remarkable doping dependence. In Na35 there is a clear low-energy peak, which turns into the dominating part of the spectrum in Na39. In contrast for slightly higher doping only a low-energy shoulder remains.
Since there is little difference in the
SC transition temperatures of Na39 and Na40 the different behavior of the low-energy SRM must be attributed to the full suppression of the AFM ordering in Na40. For Na39 magnetic
ordering as sensed by neutron diffraction sets in at 60\,K and the spin-reorientation transition is found at 44.5\,K.
In contrast neutron diffraction on Na40 does not yield any evidence for a magnetic transition.

The resonance energies can be extracted from the  data in the SC phases, see Fig. 1, and from the difference spectra of SC and NS intensities, see Fig. 3, {\red and are plotted in Fig. 4 (a)}.
The higher SRM-2 at $\sim \SI{10}{meV}$ shows a gradual doping dependency and is similarly strong as the SRM in Ba(Fe$_{0.925}$Co$_{0.075}$)$_2$As$_2$ \cite{Inosov2009}, thereby supporting a common mechanism for hole and electron doping, see Fig. 3 (a) and (b).
In contrast, the strength of SRM-1 at $E_{res,1} \sim \SI{4}{meV}$ strongly varies between the three samples peaking at the end point of AFM order.
Furthermore, SRM-1 dominates the difference excitation spectrum in Na39 where it is about twice as intense as SRM-2.
The sharp low-energy SRM-1 in Na39 forms the strongest (in absolute units) SRM reported in all iron-based superconductors studied so far.

Since INS only senses the projection of a magnetic signal perpendicular to $\bs Q$,
the comparison of the (0.5,0.5,1) and (0.5,0.5,3) spectra in the SC state indicates that the sharp extra SRM-1 is essentially polarised along the $c$ direction in Na39 \cite{note-geom}.
This conclusion is corroborated by polarized neutron scattering experiments with Na39, see supplemental materials \cite{suppl-inf}.
The higher SRM-2  lacks $c$ polarisation at an energy transfer of 10\,meV but exhibits a strong [1$\bar{1}$0]=$b$ component.
In contrast the lower SRM-1 has no $b$ contribution again in agreement with numerous observations for the lower SRM in other materials \cite{Wasser2017,Steffens2013,Zhang2014,Hu2017,Qureshi2014}.
However, SRM-1 is still essentially polarised along the $c$ direction (79(5)\% $c$ polarised and 21\% [110]=$a$ polarized) although the static magnetism is rotated from $a$ towards $c$ direction \cite{Wasser2015,Wasser2016}.
The rotation of the AFM ordered moment is thus not reflected by a rotation of the polarisation of this sharp extra low-energy SRM.
$c$ polarised extra SRMs are therefore characteristic for the SC phase of FeAs-based materials \cite{Wasser2017,Steffens2013,Zhang2014,Hu2017,Qureshi2014,Guo2019},
independently on the alignment of their ordered moments.

The exceptional strength of SRM-1 is even larger than what is suggested in Fig. 1 and 3. Because the SRM-1 is anisotropic with no component in the [1$\bar{1}$0]=$b$ direction,  the $c$ component of SRM-1 ${\chi}''_{c}\lb \bs{Q},E \rb$ should be about a factor 2 larger than the average of its components perpendicular to the scattering vector $\langle {\chi''_{\perp}}\rangle  \lb \bs{Q},E \rb$, shown in Fig. 1.

The upper and lower SRM's differ concerning their {\red $Q_L$ dependence}.
Fig.~3~(a-c) displays the intensity difference of SC state and NS for odd (Brillouin zone center) and even (zone boundary) $L$ values; data are corrected for the magnetic form factor and normalized by phonon intensities for Na35, Na39 and Na40 respectively.
SRM-1 becomes strongly suppressed at the magnetic zone boundary
{\red with the intensity following $\{1-0.88\cos(\pi\frac{L}{2})^2\}$, while
SRM-2 fully preserves its intensity. This difference suggests an enhanced
threedimensional character for SRM-1.}
The dispersion of SRM-1 is traced at intermediate positions, i.e at $ 1 \leq \text{L} \leq 2$ for Na39 and at $ 0 \leq \text{L} \leq 1$ for Na35, c.f. Fig. 3~(d) and (e) respectively.
Moreover, the resulting data points for Na35 and Na39 were combined to fit the dispersion of SRM-1 by $E(L) = \Delta + B\cdot\left|\sin \left(\frac{L-1}{2}\pi \right) \right|$, with bandwidth $B = \SI{3.2}{meV}$.
Despite their different polarisations, peak energies and dimensional character, the dispersion of SRM-1 and SRM-2 is described by the same bandwidth, which is similar to the one found in underdoped \bfca for x = \SI{4.7}{\%} \cite{Pratt2010} and which is in agreement with the universal dispersion relation $B/k_BT_c$ \cite{Lee2013}.
The {\red strong $Q_L$ dependence}
of the SRM-1 intensity is another property of this mode that is difficult to explain in the simple spin-exciton scenario.

\begin{figure}[t!]
	\begin{center}
		\includegraphics*[width=1.24\columnwidth]{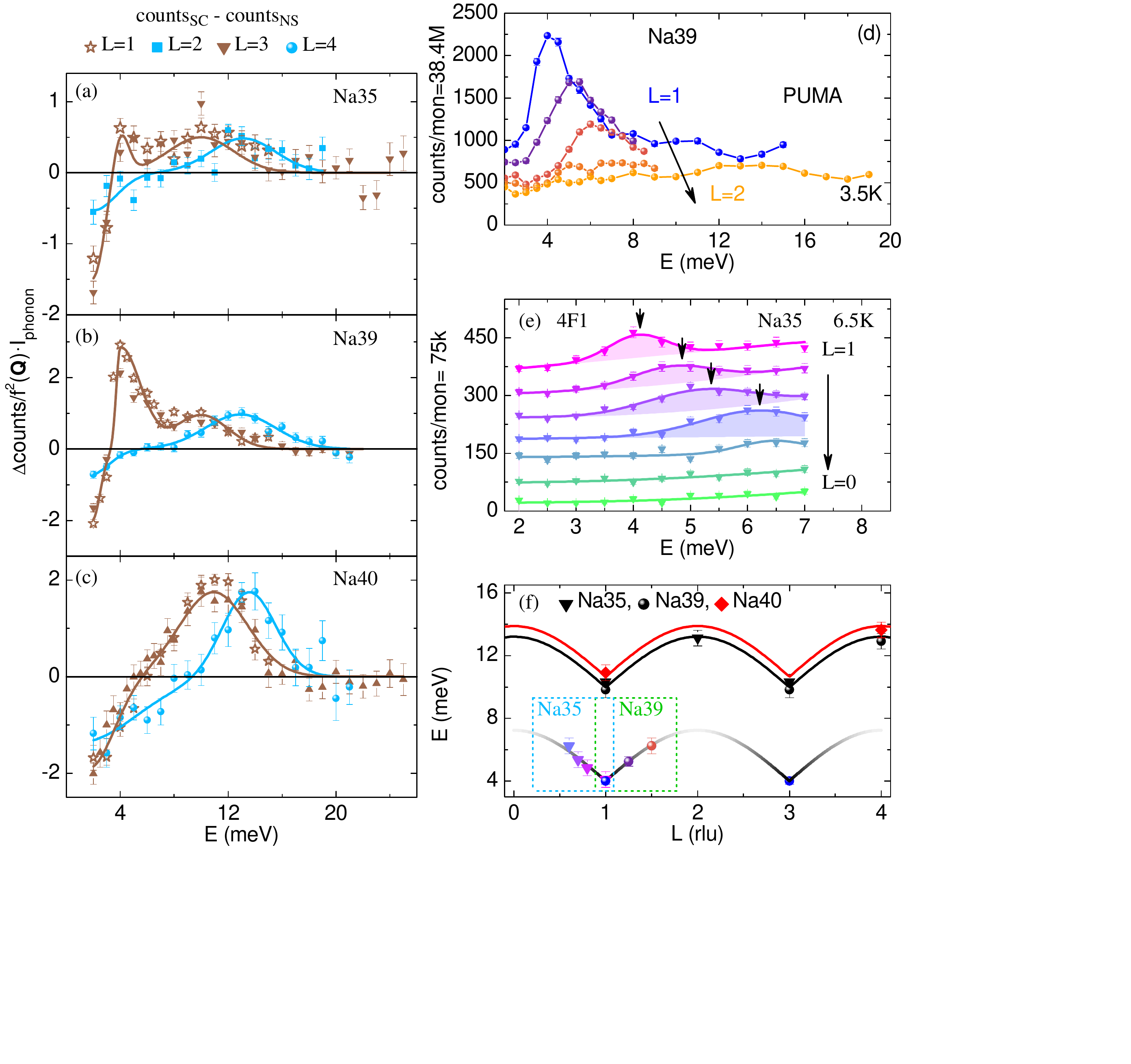}
	\end{center}
\vskip-2.3cm
	\hlcaption[Dispersion of the resonance modes.]
	{ \textbf{(a)-(c)} Normalized intensity difference of the SC state and NS at odd and even L-values for Na35, Na39 and Na40, respectively.
		Solid lines are (asymmetric, i.e. different widths above and below the peak) Gaussians peaking at different energies for odd respectively even L-values.
{\red Data in the SC state was taken at 3.5\,K, and data in the NS at 38 and 32\,K for Na35 and Na39, respectively, and at 39 and 40\,K for L=1 and 3 with Na40.}
		\textbf{(d)} Lower part of the split spin resonance mode for $1 \leq \text{L} \leq 2$ of Na39 displays a dispersion to higher energies and a reduction in intensity.	
		\textbf{(e)} Same as in (d) but for Na35 at $0 \leq \text{L} \leq 1$.
		\textbf{(f)} Fit of the dispersion as described in the main text in which the data points of Na35 and Na39 were combined.
		Disregarding the color, all triangular symbols belong to Na35, while all spherical symbols belong to Na39.
	}
	\label{fig:disp}
\end{figure}

\begin{figure}[t]
	\includegraphics*[width=1.2\columnwidth]{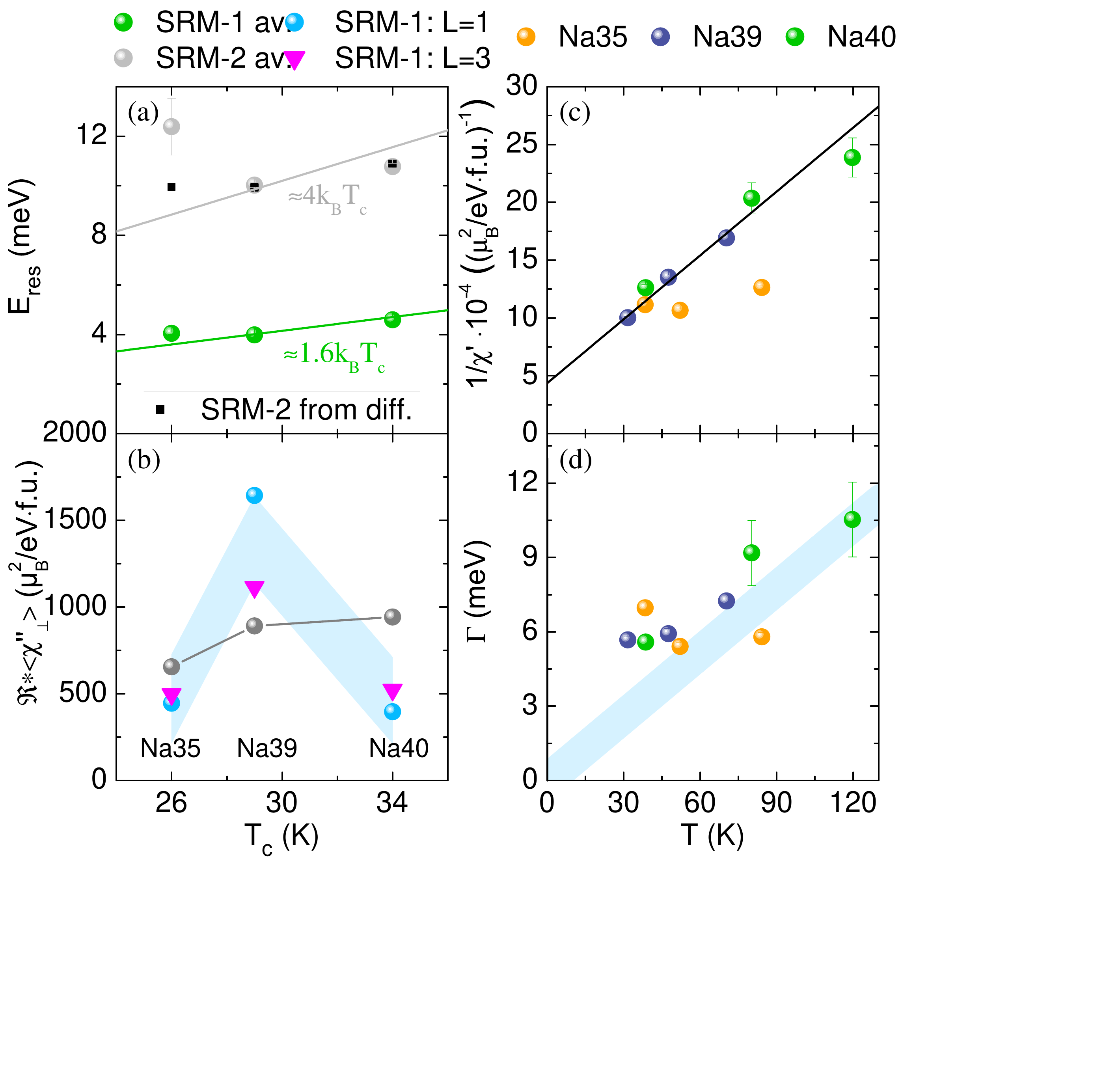} 
\vskip-2cm
	\hlcaption[Energies and amplitudes of magnetic scattering.]
	{ 
Parameters describing the magnetic response obtained from the spectra and difference spectra. (a) and (b) show the energies and the amplitudes of the SRM-1 and SRM-2 log-normal functions plotted against the \tc \ of the three samples studied. For the high-energy SRM-2 we include the value obtained from fitting the intensity differences shown in {\red Fig. 3 (black squares)}. (c) and (d) resume the single-relaxor parameters describing the normal-state response traced against the temperature. One recognizes that Na39 and Na40 behave very similarly concerning both the amplitude and the characteristic energies, while the response of Na35 is determined by the stronger static AFM order. The linear fit in (c) corresponds to a Curie-Weiss law, and the blue bar in (d) denotes the Planck scaling $\Gamma = k_B T$.
	}
	\label{fig:pd}
\end{figure}

The spectra in the SC state exhibit a spin gap, {\red which can be } described by the log-normal functions. In contrast all NS spectra can be reasonably well described by the single relaxor formula $\chi''\lb \bs{Q},E \rb = \chi'\lb \bs{Q},0 \rb \frac{\Gamma E}{\Gamma ^2 +E^2}$, see Fig. 2, that is expected for a paramagnetic itinerant magnetic system.
This might appear astonishing, as Na35 and Na39 enter the SC state from an AFM ordered phase with essentially $c$-aligned magnetic moments. For an ordered material
one may expect spin gaps arising from anisotropy, as it was reported for the parent compound \cite{QureshiBraden2012,Wang2013}. However, in Na35 and Na39
the ordered moments are heavily reduced with respect to undoped BaFe$_2$As$_2$.
In contrast to electron-doped \bfcaud \cite{Wasser2017} the AFM
order in these hole-doped compounds appears much softer; there is no evidence for a gap in any of the channels in the normal conducting AFM phases. In Na39 just above \tc there is little evidence for an anisotropy
between $c$ and $a$ polarised excitations, see Fig. 2 (d), while in Na35 the $c$ axis response is slightly suppressed, see Fig. 2 (a), as it is expected for AFM ordering with moments along $c$, and as it was also reported for very underdoped Sr$_{1-x}$Na$_x$Fe$_2$As$_2$ only \cite{Guo2019}.
Again this suggests that the AFM ordered moment  and its directional pinning in Na39 is too small to significantly impact the inelastic magnetic response, see also below.
The difference between the two stronger doped compounds
and Na35 is also seen when tracing the single-relaxor parameters against temperature in the NS, see Fig. 4 (c,d). For the inverse of the amplitude, $\frac{1}{\chi'\lb \bs{Q},0 \rb}$, a linear temperature dependency corresponding a Curie-Weiss law, $ \chi'\lb \bs{Q},0 \rb \propto \frac{1}{T-\Theta}$, can be expected \cite{steffens2011}, which indeed is roughly fulfilled in both compounds with identical slope and  Weiss temperature $\Theta$=22\,K. In contrast the AFM ordering in Na35 results in qualitatively different temperature dependencies.
Also the plot of the characteristic energy, $\Gamma$, indicates the different behavior of Na35 and Na39/Na40. The latter two samples show a similar temperature dependency that most interestingly is very close to the Planck scaling for the characteristic energy, $\Gamma = k_B T$ \cite{Zanen}, indicated by the blue bar.

\begin{figure*}[t]
\begin{center}
	\includegraphics*[width=\textwidth]{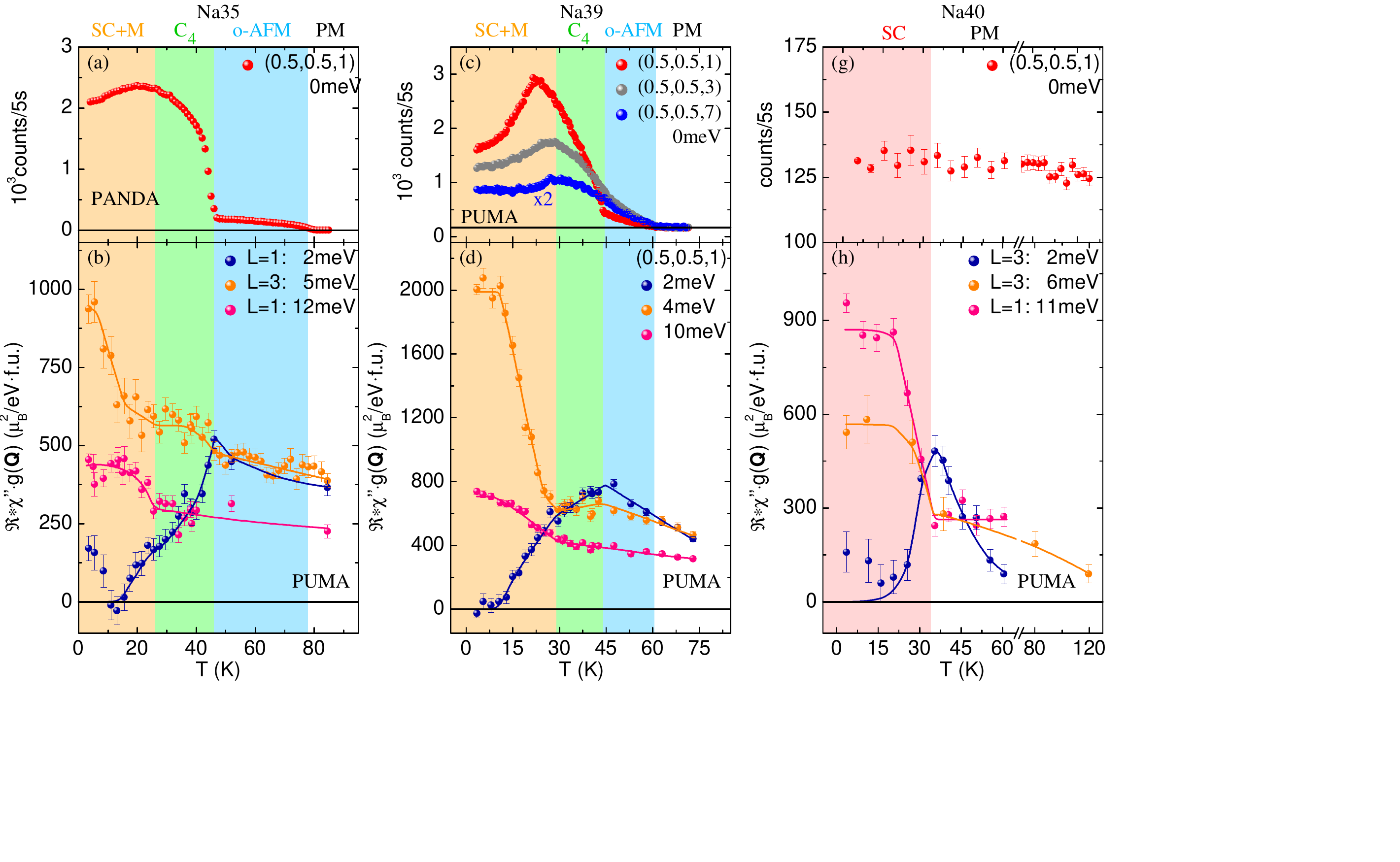} 
\end{center}
\vskip-2.5cm
	\hlcaption[Temperature dependence of the elastic and inelastic response.]
	{ \textbf{(a)} The temperature dependent scattering of the \hkl(0.5 0.5 1) magnetic Bragg peak of Na35 shows the three phase transitions, data taken from \cite{Wasser2015} and binned within \SI{1}{K}.
		\textbf{(b)} Inelastic response of Na35 displays a peak for \SI{2}{meV} at \treo, while the intensity uptake at \SI{5}{meV} and \SI{12}{meV} corresponds to the formation of the resonance mode.
		\textbf{(c)} Elastic response of Na39 determined at the \hkl(0.5 0.5 L) positions that sense $c$ and $a$ polarized moments differently \cite{note-geom}, data taken from \cite{Wasser2016}.
		\textbf{(d)} Inelastic response of Na39 at \hkl(0.5 0.5 1) showing for \SI{2}{meV} a maximum at \treo  ~ and a suppression in the SC phase.
The emergence of the SRM's result in increasing intensities at \SI{4}{meV} and \SI{10}{meV} below \tc .
	    \textbf{(e)} The constant elastic scattering at the \hkl(0.5 0.5 1) magnetic Bragg-position for Na40 shows the absence of magnetic ordering down to the lowest temperatures, data taken from \cite{Wasser2016}.
		\textbf{(f)} Inelastic response of Na40 showing a maximum for \SI{2}{meV} at \tc \, and an intensity uptake due to the SRM formation at \SI{6}{meV} and \SI{11}{meV}, respectively.
		All lines are guides to the eye and the colored backgrounds correspond to the SC, C$_4$ and o-AFM phases. 	
}
	\label{fig:tdep}
\end{figure*}

Further insight into the character of the SRMs can be obtained by comparing the temperature dependencies of the inelastic signals with those
of the elastic Bragg intensities for the three compositions, see Fig.~\ftdep.
The elastic signals measured on the same crystals were taken from references \cite{Wasser2015,Wasser2016}.
For all three materials there is a clear suppression of the magnetic signal at 2\,meV in the SC phase resulting from the opening of the gap.
While in Na40 this 2\,meV signal is the strongest just above $T_c$ it is already suppressed through the AFM ordering in Na35 and Na39.
At (0.5 0.5 1) the 2\,meV signal essentially corresponds to $c$ polarised magnetic excitations, which are little influenced by the
upper magnetic transition into the o-AFM phase with moments aligned in-plane. The fact that the signal further increases upon cooling in the
AFM state corroborates the softer character of the magnetism in these hole-doped materials compared to electron doping \cite{Wasser2017}.
The maximum 2\,meV signal is observed just above the
reorientation transition. A continuous rotation of the moments would even require complete softening and divergence of this susceptibility. In the C$_4$
phase the $c$ polarised excitations become the longitudinal ones, for which strong suppression is expected.
{\red Note, however, that in the double-q magnetic structure with moments pointing along $c$, half of the Fe sites exhibit zero magnetic moment,
which can result in strong $c$-polarized fluctuations.}

The magnetic signals near the energies of the two SRMs spontaneously increase in the SC phase.
The most remarkable upturn is observed for the low-energy signals in the two materials with coexisting AFM order, Na35 and Na39, underlining
that SRM-1 emerges with the opening of the SC gap.
This enormous increase of
inelastic intensity is accompanied by the partial suppression of the magnetic Bragg intensity and thus by the suppression of ordered magnetic moment described
in reference \cite{Wasser2016}. The by far strongest increase of SRM scattering in the SC state is found in Na39 where it is accompanied by the most
pronounced drop of static intensity suggesting that the SRM is closely related with the suppressed magnetic moment, as it will be quantitatively analyzed below.

\section*{Discussion}

The comparison of the three samples with little variation of the Na doping gives a detailed insight in the character and origin of the
split SRMs appearing in various FeAs-based superconductors.
Compared to the electron doped phase diagrams the AFM ordering in the hole-doped series seems to be softer. While the anisotropy gaps in the
electron-doped compounds are sizeable and thus effectively pin the ordered moment to lie parallel to the FeAs layers \cite{Wasser2017}, no such gaps are
visible in the magnetic response of the two Na doped BaFe$_2$As$_2$ compounds that exhibit magnetic order and the spin-reorientation transition, Na35 and Na39.
For hole doping at temperatures just above the SC transition, there is thus much more spectral weight situated at low energies, which might
be the clue to understand the higher SC transition temperatures in hole versus electron doped compounds.

In spite of the spin-reorientation transition the magnetic excitations in SC Ba$_{1-x}$Na$_x$Fe$_2$As$_2$, i.e. the SRMs, qualitatively resemble those in
electron doped compounds. There is a clear superposition of at least two SRMs with similar energy dispersion along L, from which the upper possesses a two-dimensional and
the lower one a three-dimensional character. Most remarkably the lower SRM-1 exhibits qualitatively the same
anisotropy with a predominant $c$ polarized character as previously observed in materials with $a$ aligned magnetism. The anisotropy of the SRM-1 thus does not follow the spin reorientation, but rather a predominantly $c$ polarized low-energy SRM is ubiquitous in FeAs-based superconductors.

There is some ongoing debate whether the resonance excitations can also be explained within the $s^{++}$ orbital fluctuations scenario. However, the low energy
of only $\sim$4\,meV for the lower SRM-1, which without any doubt emerges in the SC state, rules out such an explanation for this mode. Its energy
is clearly below twice the sum of a smaller and a larger gap for all three materials studied, which was formulated as a strict criterion for $s{\pm}$ pairing by Korshunov et al. \cite{Korshunov2016}.

\begin{table}
\caption{ Quantitative comparison of the static and dynamic signals for the three compounds Na35, Na39 and Na40 calculated
by integrating the static and inelastic neutron scattering response, {\red for precise temperatures see caption of Fig. 3}. All quantities are given in Bohr magnetons squared,
$\mu_B^2$ and fluctuating moments correspond to the sum of the directions. $\mu ^2_\textrm{SRM-1}$ and $\mu ^2_\textrm{SRM-2}$
give the total squared moment in the two resonance modes integrated up to 20\,meV; $\mu ^2_\textrm{NS}$ the fluctuating moments
in the NS;    $\Delta\mu ^2_{stat}$  is the reduction of ordered moment squared upon entering the SC state.
}
\begin{tabular}{l  c  c  c  }
\hline \hline
  $\mu ^2[\mu_B^2]$   &   \; Na35  \;   &  \;  Na39  \;  &  \;  Na40 \;   \\
\hline
 $\mu ^2_\textrm{SRM-1}$  & 0.004 & 0.012 & 0.003  \\
 $\mu ^2_\textrm{SRM-2}$  & 0.065 & 0.073 & 0.070  \\
\hline
 $\mu ^2_\textrm{NS}$     & 0.056 & 0.070 & 0.052  \\
 \hline
  $\Delta\mu ^2_{stat}$     & -0.006 & -0.012 & /  \\

\end{tabular}
\end{table}

The soft AFM ordering {\red and the stronger coupling between SC and the $c$-aligned order seem } to be the cause of the exceptionally strong SRM-1 appearing for the Na content closest to the doping-induced full suppression of magnetic order.
This strong inelastic mode, whose peak amplitude is the strongest one reported for any FeAs-based SC so far, is accompanied by the pronounced suppression of the
ordered $c$-polarized moment upon entering the SC state.
In order to quantify this relation between inelastic and elastic signals, the energy and $Q$-integrated strength of  SRM-1 as well as those of SRM-2 and of the
 NS response can be calculated from the
signal in absolute units. Since we are interested in temperature differences and since there seems to be no impact of SC at high
energies, {\red we average the generalized susceptibility over $\bs{Q}$ in a Brillouin zone and integrate it over energy from 0 to 20\,meV, see methods section and \cite{suppl-inf}.
These calculations take the geometry factors of the signals and the $\bs{Q}_L$ dispersion of the amplitudes and
peak energies of the two SRM's into account,} and the results are resumed in table I.

Although the lower SRM-1 exhibits by far the largest
peak amplitude its total fluctuating moment is much smaller than that of SRM-2, because it looses rapidly its weight when $L$ varies away from the odd zone-center values, see Fig. 3, {\red in addition it essentially concerns only one spin direction while SRM-2 is isotropic. Furthermore, in the
description of the SC response by two log-normal functions the upper one contains also the part of the spectrum that does not
change upon entering the SC state yielding its much larger total weight. These fluctuating moments
can be compared to the reduction of the ordered moment squared obtained from the magnetic Bragg scattering, see  \cite{suppl-inf}.
For Na35 the ordered moment decreases from 0.222 to 0.208 $\mu_B$ and for Na39 from 0.189 to 0.155, which corresponds
to a sizeable reduction of the square of the ordered moment, see table I.
Indeed the total fluctuating moment of SRM-1 quantitatively follows the reduction of the ordered moment squared.
There is also a clear total enhancement of magnetic fluctuations in the SC state, which points to a significant gain in exchange energy
similar to other unconventional superconductors \cite{Scalapino,note-exch}.}

The good agreement between the suppressed magnetic order and the strength of SRM-1 for the two compounds with coexistence of SC and AFM order
suggests that the SRM-1 is related to the suppression of magnetic order. The fact that SRM-1 is essentially $c$ polarized nicely agrees with the
observation that $c$-ordered magnetism competes more strongly with SC or becomes more strongly suppressed \cite{Boehmer2015,Wasser2016}.
For Na40, which does not exhibit any AFM ordering, the weight of SRM-1 can stem from the suppression of anisotropic low-energy fluctuations
in the normal and paramagnetic state, see Fig. 2.
In view of the observation of a qualitatively similar mode at almost the same energy of $\sim$4\,meV in so many different compounds \cite{Steffens2013,Zhang2014,Zhang2013b,Qureshi2014} there must be an intrinsic {\red cause} for it
basing on the AFM or nearly AFM band structure. The spin reorientation seems not to have a qualitative impact on this mode but the $c$ aligned magnetic order couples more strongly with SC and thus yields a stronger SRM-1.

The coexistence of AFM order and SC has been too little studied in theory, just for the case of a single band and without spin-orbit coupling it was found
that {\red transverse} excitations should remain ungapped and that the longitudinal channel develops a resonance \cite{Ismer2007,Rowe2012}. This longitudinal resonance might be related to the peculiar behavior of SRM-1.

In summary there are three properties of the lower SRM in Na-doped BaFe$_2$As$_2$ that challenge the mostly accepted exciton picture.
Firstly the split SRMs partially share the same polarization while there should only exist a single bound state below the
electron-hole continuum. Secondly, the {\red amplitude of the lower SRM exhibits a strong $Q_L$ dependence} in contrast to the highly two-dimensional electron band structure well reflected by the high energy SRM. And thirdly, the lower SRM sharply peaks at the suppression of the AFM ordering, while the exciton intensity should be
determined essentially by the distance to the continuum, which in Na-doped BaFe$_2$As$_2$ cannot vary so sharply.
Furthermore this exceptionally strong SRM-1 in Na39 exhibits the same polarization as previously observed in electron doped systems in spite of the
spin reorientation. The stronger competition between $c$ aligned magnetic order and SC in this hole-doped material seems to imply the strong low-energy
anisotropic SRM.

\section*{Methods}

\bnfax single crystals were synthesised by a self-flux method as described in Ref. \cite{Aswartham2012}.
Thereby we obtained samples with x = 0.35 (Na35) and $m = \SI{294}{mg}$, x = 0.39 (Na39) and $m = \SI{172}{mg}$ and x = 0.40 (Na40) and $m = \SI{42}{mg}$.
The high sample quality was assessed by EDX, magnetisation and transport measurements \cite{Aswartham2012} as well as neutron scattering \cite{Wasser2015,Wasser2016}.
Note that there is some discrepancy between various phase diagrams reported for  \bnfax \, \cite{Wasser2016}; here we label the crystals according to their Na content
as determined by the EDX measurements.
The single crystals were air-tightly sealed in thin Al cans containing Ar atmosphere in order to prevent potential degradation in air.
Unpolarised INS experiments were performed at the thermal triple-axis-spectrometer (TAS) PUMA at Meier-Leibnitz Zentrum in Garching, Germany, and at the cold TAS 4F1 at the Laboratoire L\'eon Brillouin  in Saclay, France.
Polarised INS experiments were conducted at the thermal TAS IN22 at Institut Laue Langevin in Grenoble, France, where polarised neutrons were generated and analysed via a Heusler-monochromator and -analysor, respectively, see the supplemental material \cite{suppl-inf}.
Each instrument was operated in constant-$k_f$ mode and $k_f$ was fixed to \SI{2.662}{\reza} in conjunction with a pyrolytic graphite (PG) filter in case of thermal TAS and to \SI{1.55}{\reza} in conjunction with a cooled beryllium filter in case of cold TAS.

The samples were aligned within the \hkl[110]/\hkl[001] scattering geometry corresponding to orthorhombic $a$ and $c$  directions.
Throughout this paper we use the tetragonal notation and denote the scattering vector $\bs{Q}$ in reciprocal lattice units.
Whenever possible, the background was determined by rotating the sample, fitted and subtracted from the data, see supplemental material \cite{suppl-inf}. Furthermore, corrections for the thermal population (Bose factor) and the energy dependent fraction of higher-order neutrons (monitor factor) were applied. For consistency two symmetry-equivalent \q positions at $ \hkl(0.5 0.5 1)$ and $\hkl(0.5 0.5 3)$ were probed by series of E-scans (constant \q) and \q-scans (constant E).

The intensities in INS correspond to the double-differential cross section, $\frac{d^2\sigma}{d\Omega dE}$, scaled with the flux and with the transmission factor and {\red convoluted} with the resolution function of the spectrometer. We followed the approach described in reference \cite{Qureshi2014a} to convert the energy spectra in Fig. \fescan to an absolute scale by normalizing through phonon scattering \cite{suppl-inf} and estimating the resolution of PUMA spectrometer by the  \reslib-code \cite{Zheludev2006}.
Note, however, that the folding with the resolution function is nontrivial so that this normalized INS signal corresponds to the double-differential cross section only when the resolution is better than the typical variation of the cross section with the scattering vector, $\bs Q$, or energy transfer, $E$. This condition is not always fulfilled for the magnetic response in FeAs-based SC's.
The double differential cross section is given by the imaginary part of the generalized dynamic susceptibility:
\begin{align}
\begin{split}
\frac{d^2\sigma}{d \Omega dE} &= \left(\frac{\gamma r_0}{2} \right) ^2 \cdot \frac{k_f}{k_i} \cdot \frac{N}{ \pi \mu_B^2} \cdot \frac{\e^{-2W(\boldsymbol{Q})}}{1-\e^{-\frac{E}{k_BT}}}  \times\\
& \qquad f^2(\boldsymbol{Q} ) \cdot \chi''(\boldsymbol{Q}, E)
 \times \\
&\qquad \sum_{\alpha, \beta} \left( \delta_{\alpha \beta} - \frac{\boldsymbol{Q}_{\alpha} \cdot \boldsymbol{Q}_{\beta}}{|\boldsymbol{Q}|^2} \right),
\end{split}
\label{eq:ddcs}
\end{align}
where $\sum_{\alpha, \beta} \left( \delta_{\alpha \beta} - \boldsymbol{Q}_{\alpha} \cdot \boldsymbol{Q}_{\beta}/|\boldsymbol{Q}|^2 \right)$ corresponds to a factor two for isotropic magnetic signals and to the average of the two components perpendicular to $\bs Q$ in an anisotropic case.
The data were corrected for this geometry factor yielding  the average
of the two perpendicular components of  $\chi''\lb \bs{Q},E \rb$,   $\langle {\chi''_{\perp}}\rangle  \lb \bs{Q},E \rb$ . The fact that the signal still contains the resolution folding is reflected by
the label $\Re \star \langle {\chi''_{\perp}}\rangle  \lb \bs{Q},E \rb$ in Figures 1, 2 and 5.

The lines in Figures 1 and 2 are a fit of a model of $\chi''(\boldsymbol{Q}, E)$ convoluted with the instrumental resolution.
For the split SRMs a phenomenological model of two ($j = 1,2$) log-normal functions
$$
L_j(E) = \frac{A_j}{\sqrt{(2\pi)}w_j E} \exp\left({-\frac{1}{2} \left(\frac{\ln(E_{res,j})-\ln(E)}{w_j}\right)^2}\right)
$$
was used, that were previously shown to well describe SC magnetic excitations in Fe-based SC's \cite{Tucker2014a,Wang2016c,Wasser2017}.
In contrast the NS response is described by a single relaxor function $\chi''\lb \bs{Q},E \rb = \chi'\lb \bs{Q},0 \rb \cdot \frac{\Gamma \cdot E}{\Gamma^2 + E^2}$.
{\red For the calculation of the fluctuating moments the $Q$ dependence was described by $exp(-0.5(\frac{h-0.5}{w_{hk}})^2)exp(-0.5(\frac{k-0.5}{w_{hk}})^2)$ with a constant width that corresponds to
a half width at half maximum of 0.027 reciprocal lattice units in $(hhL)$ scans, see also \cite{suppl-inf}. The dispersion of the resonance energies as well as that of the amplitude
of SRM-1 was described by the formulas given in the RESULTS section. The parameters shown in Fig. 4 (averaged for L=1 and 3) were used
and the width parameters of SRM-1(SRM-2), $w_{1,2}$, are 0.151(0.670), 0.151(0.532), 0.151(0.330) for Na35, Na39 and Na40, respectively.
For the anisotropic SRM-1 a geometry factor of 1.5 was applied, while for the isotropic signals the three directions were summed up.}
Note that error bars in all figures correspond to the standard deviation of the intensity and do not include errors of the normalization process, which would be in the range of $\sim$\SI{20}{\%} \cite{Xu2013}.

\section*{DATA AVAILABILITY}

All relevant data that support the findings of this study are available from the corresponding author on request.

\section*{Acknowledgements}

This work was supported by the Deutsche Forschungsgemeinschaft (DFG) through the Priority Programme SPP1458 (BE1749/13, BU887/15-1 and BR2211/1-1).
S. W. thanks the DFG for funding in the Emmy Noether programme (WU595/3-3).

\section*{Competing Interests}
The authors declare no competing interests.

\section*{Contributions}
F. W., J.-T. P., Y. S., P. S., K. S. and M. B. performed the experiments; S. W. and S. A. synthesised the samples; F. W. and M. B. analysed the data; F. W. and M. B. wrote the manuscript with contributions from J.-T. P., Y. S., P. S, K. S., S. W., A. S., and B. B.; F. W., B. B. and M. B. organized the work.


\begin{thebibliography}{10}

\bibitem{Kamihara2008}
Kamihara, Y., Watanabe, T., Hirano, M. \& Hosono, H., Iron-Based Layered
  Superconductor La(O$_{1-x}$F$_x$)FeAs (x = 0.05-0.12) with T$_c$ = 26 K.
  \textit{J. Am. Chem. Soc.} \textbf{130}, 3296 (2008).

\bibitem{Dai2015a}
Dai, P., Antiferromagnetic order and spin dynamics in iron-based
  superconductors. \textit{Rev. Mod. Phys.} \textbf{87}, 855 (2015).

\bibitem{Hirschfeld2011}
Hirschfeld, P.~J., Korshunov, M.~M. \& Mazin, I.~I., Gap symmetry and structure
  of Fe-based supersonductors. \textit{Rep. Prog. Phys.} \textbf{74}, 124508
  (2011).

\bibitem{Korshunov2008}
Korshunov, M.~M. \& Eremin, I., Theory of magnetic excitations in iron-based
  layered superconductors. \textit{Phys. Rev. B} \textbf{78}, 140509 (2008).

\bibitem{Korshunov2016}
Korshunov, M.~M., Shestakov, V.~A. \& Togushova, Y.~N., Spin resonance peak in
  Fe-based superconductors with unequal gaps. \textit{Phys. Rev. B}
  \textbf{94}, 094517 (2016).

\bibitem{Christianson2008}
Christianson, A.~D., Goremychkin, E.~A., Osborn, R., Rosenkranz, S., Lumsden,
  M.~D., Malliakas, C.~D., Todorov, I.~S., Claus, H., Chung, D.~Y., Kanatzidis,
  M.~G., Bewley, R.~I. \& Guidi, T., Unconventional superconductivity in
  Ba$_{0.6}$K$_{0.4}$Fe$_2$As$_2$ from inelastic neutron scattering.
  \textit{Nature} \textbf{456}, 930 (2008).

\bibitem{Inosov2009}
Inosov, D.~S., Park, J.~T., Bourges, P., Sun, D.~L., Sidis, Y., Schneidewind,
  A., Hradil, K., Haug, D., Lin, C.~T., Keimer, B. \& Hinkov, V., Normal-state
  spin dynamics and temperature-dependent spin-resonance energy in optimally
  doped BaFe$_{1.85}$Co$_{0.15}$As$_2$. \textit{Nat. Phys.} \textbf{6}, 178
  (2009).

\bibitem{Zhang2011}
Zhang, C., Wang, M., Luo, H., Wang, M., Liu, M., Zhao, J., Abernathy, D.~L.,
  Maier, T.~A., Marty, K., Lumsden, M.~D., Chi, S., Chang, S.,
  Rodriguez-Rivera, J.~A., Lynn, J.~W., Xiang, T., Hu, J. \& Dai, P., Neutron
  Scattering Studies of spin excitations in hole-doped
  \ce{Ba_{0.67}K_{0.33}Fe2As2} superconductor. \textit{Scientific Reports}
  \textbf{1}, 115 (2011).

\bibitem{Lee2013}
Lee, C.~H., Steffens, P., Qureshi, N., Nakajima, M., Kihou, K., Iyo, A.,
  Eisaki, H. \& Braden, M., Universality of the Dispersive Spin-Resonance Mode
  in Superconducting BaFe$_2$As$_2$. \textit{Phys. Rev. Lett.} \textbf{111},
  167002 (2013).

\bibitem{Qureshi2012}
Qureshi, N., Steffens, P., Drees, Y., Komarek, A.~C., Lamago, D., Sidis, Y.,
  Harnagea, L., Grafe, H.-J., Wurmehl, S. \& andM. Braden, B.~B., Inelastic
  Neutron-Scattering Measurements of Incommensurate Magnetic Excitations on
  Superconducting LiFeAs Single Crystals. \textit{Phys. Rev. Lett.}
  \textbf{108}, 117001 (2012).

\bibitem{Wang2016f}
Wang, Q., Shen, Y., Pan, B., Hao, Y., Ma, M., Zhou, F., Steffens, P., Schmalzl,
  K., Forrest, T.~R., Abdel-Hafiez, M., Chen, X., Chareev, D.~A., Vasiliev,
  A.~N., Bourges, P., Sidis, Y., Cao, H. \& Zhao, J., Strong interplay between
  stripe spin fluctuations, nematicity and superconductivity in FeSe.
  \textit{Nat. Mater.} \textbf{15}, 159 (2016).

\bibitem{Steffens2013}
Steffens, P., Lee, C.~H., Qureshi, N., Kihou, K., Iyo, A., Eisaki, H. \&
  Braden, M., Splitting of Resonance Excitations in Nearly Optimally Doped
  \ce{Ba(Fe_{0.94}Co_{0.06})_2As_2}: An Inelastic Neutron Scattering Study with
  Polarization Analysis. \textit{Phys. Rev. Lett.} \textbf{110}, 137001 (2013).

\bibitem{Wasser2017}
Wa\ss{}er, F., Lee, C.~H., Kihou, K., Steffens, P., Schmalzl, K., Qureshi, N.
  \& Braden, M., Anisotropic resonance modes emerging in an antiferromagnetic
  superconducting state. \textit{Scientific Reports} \textbf{7}, 10307 (2017).

\bibitem{Zhang2014}
Zhang, C., Song, Y., Regnault, L.-P., Su, Y., Enderle, M., Kulda, J., Tan, G.,
  Sims, Z.~C., Egami, T., Si, Q. \& Dai, P., Anisotropic neutron spin resonance
  in underdoped superconducting NaFe$_{1-x}$Co$_x$As. \textit{Phys. Rev. B}
  \textbf{90}, 140502 (2014).

\bibitem{Song2017}
Song, Y., Wang, W., Zhang, C., Gu, Y., Lu, X., Tan, G., Su, Y., Bourdarot, F.,
  Christianson, A.~D., Li, S. \& Dai, P., Temperature and polarization
  dependence of low-energy magnetic fluctuations in nearly-optimal-doped
  NaFe$_{0.9785}$Co$_{0.0215}$As. \textit{arXiv: cond-mat} 1710.11083 (2017).

\bibitem{Hu2017}
Hu, D., Zhang, W., Wei, Y., Roessli, B., Skoulatos, M., Regnault, L.~P., Chen,
  G., Song, Y., Luo, H., Li, S. \& Dai, P., Spin excitation anisotropy in
  optimal-isovalent-doped superconductor BaFe$_2$(As$_{0.7}$P$_{0.3}$)$_2$.
  \textit{arXiv: cond-mat} 1711.01497 (2017).

\bibitem{Luo2013}
Luo, H., Wang, M., Zhang, C., Lu, X., Regnault, L.-P., Zhang, R., Li, S., Hu,
  J. \& Dai, P., Spin Excitation Anisotropy as a Probe of Orbital Ordering in
  the Paramagnetic Tetragonal Phase of Superconducting
  \ce{BaFe_{1.904}Ni_{0.096}As2}. \textit{Phys. Rev. Lett.} \textbf{111},
  107006 (2013).

\bibitem{Zhang2013b}
Zhang, C., Liu, M., Su, Y., Regnault, L.-P., Wang, M., Tan, G., Br\"uckel, T.,
  Egami, T. \& Dai, P., Magnetic anisotropy in hole-doped superconducting
  \ce{Ba_{0.67}K_{0.33}Fe2As2} probed by polarized inelastic neutron
  scattering. \textit{Phys. Rev. B} \textbf{87}, 081101 (2013).

\bibitem{Qureshi2014}
Qureshi, N., Lee, C.~H., Kihou, K., Schmalzl, K., Steffens, P. \& Braden, M.,
  Anisotropy of incommensurate magnetic excitations in slightly overdoped
  ${\mathrm{Ba}}_{0.5}{\mathrm{K}}_{0.5}{\mathrm{Fe}}_{2}{\mathrm{As}}_{2}$
  probed by polarized inelastic neutron scattering experiments. \textit{Phys.
  Rev. B} \textbf{90}, 100502 (2014).

\bibitem{Knolle2011}
Knolle, J., Eremin, I., Schmalian, J. \& Moessner, R., Magnetic resonance from
  the interplay of frustration and superconductivity. \textit{Phys. Rev. B}
  \textbf{84}, 180510 (2011).

\bibitem{Lv2014}
Lv, W., Moreo, A. \& Dagotto, E., Double magnetic resonance and spin anisotropy
  in Fe-based superconductors due to static and fluctuating antiferromagnetic
  orders. \textit{Phys. Rev. B} \textbf{89}, 104510 (2014).

\bibitem{Wang2016c}
Wang, M., Yi, M., Sun, H.~L., Valdivia, P., Kim, M.~G., Xu, Z.~J., Berlijn, T.,
  Christianson, A.~D., Chi, S., Hashimoto, M., Lu, D.~H., Li, X.~D.,
  Bourret-Courchesne, E., Dai, P., Lee, D.~H., Maier, T.~A. \& Birgeneau,
  R.~J., Experimental elucidation of the origin of the `double spin resonances'
  in
  $\mathrm{Ba}{({\mathrm{Fe}}_{1\ensuremath{-}x}{\mathrm{Co}}_{x})}_{2}{\text{As}}_{2}$.
  \textit{Phys. Rev. B} \textbf{93}, 205149 (2016).

\bibitem{Yu2014}
Yu, R., Zhu, J.-X. \& Si, Q., Orbital-selective superconductivity, gap
  anisotropy, and spin resonance excitations in a multiorbital $t-J_1-J_2$
  model for iron pnictides. \textit{Phys. Rev. B} \textbf{89}, 024509 (2014).

\bibitem{QureshiBraden2012}
Qureshi, N., Steffens, P., Wurmehl, S., Aswartham, S., B\"uchner, B. \& Braden,
  M., Local magnetic anisotropy in \ce{BaFe_2As_2}: A polarized inelastic
  neutron scattering study. \textit{Phys. Rev. B} \textbf{86}, 060410 (2012).

\bibitem{Wang2013}
Wang, C., Zhang, R., Wang, F., Luo, H., Regnault, L.~P., Dai, P. \& Li, Y.,
  Longitudinal Spin Excitations and Magnetic Anisotropy in
  Antiferromagnetically Ordered \ce{BaFe2As2}. \textit{Phys. Rev. X}
  \textbf{3}, 041036 (2013).

\bibitem{Song2013}
Song, Y., Regnault, L.-P., Zhang, C., Tan, G., Carr, S.~V., Chi, S.,
  Christianson, A.~D., Xiang, T. \& Dai, P., In-plane spin excitation
  anisotropy in the paramagnetic state of NaFeAs. \textit{Phys. Rev. B}
  \textbf{88}, 134512 (2013).

\bibitem{Borisenko2016}
Borisenko, S.~V., Evtushinsky, D.~V., Liu, Z.-H., Morozov, I., Kappenberger,
  R., Wurmehl, S., Buchner, B., Yaresko, A.~N., Kim, T.~K., Hoesch, M., Wolf,
  T. \& Zhigadlo, N.~D., Direct observation of spin-orbit coupling in
  iron-based superconductors. \textit{Nat. Phys.} \textbf{12}, 311 (2016).

\bibitem{Avci2014}
Avci, S., Chmaissem, O., Allred, J., Rosenkranz, S., Eremin, I., Chubukov, A.,
  Bugaris, D., Chung, D., Kanatzidis, M., Castellan, J.-P., Schlueter, J.,
  Claus, H., Khalyavin, D., Manuel, P., Daoud-Aladine, A. \& Osborn, R.,
  Magnetically driven suppression of nematic order in an iron-based
  superconductor. \textit{Nat Commun} \textbf{5}, 3845 (2014).

\bibitem{Wasser2015}
Wa\ss{}er, F., Schneidewind, A., Sidis, Y., Wurmehl, S., Aswartham, S.,
  B\"uchner, B. \& Braden, M., Spin reorientation in
  ${\mathrm{Ba}}_{0.65}{\mathrm{Na}}_{0.35}{\mathrm{Fe}}_{2}{\mathrm{As}}_{2}$
  studied by single-crystal neutron diffraction. \textit{Phys. Rev. B}
  \textbf{91}, 060505 (2015).

\bibitem{Boehmer2015}
B\"ohmer, A.~E., Hardy, F., Wang, L., Wolf, T., Schweiss, P. \& Meingast, C.,
  Superconductivity-induced re-entrance of the orthorhombic distortion in
  Ba$_{1-x}$K$_x$Fe$_2$As$_2$. \textit{Nat. Commun.} \textbf{6}, 7911 (2015).

\bibitem{Taddei2016a}
Taddei, K.~M., Allred, J.~M., Bugaris, D.~E., Lapidus, S., Krogstad, M.~J.,
  Stadel, R., Claus, H., Chung, D.~Y., Kanatzidis, M.~G., Rosenkranz, S.,
  Osborn, R. \& Chmaissem, O., Detailed magnetic and structural analysis
  mapping a robust magnetic ${C}_{4}$ dome in
  ${\mathrm{Sr}}_{1\ensuremath{-}x}{\mathrm{Na}}_{x}{\mathrm{Fe}}_{2}{\mathrm{As}}_{2}$.
  \textit{Phys. Rev. B} \textbf{93}, 134510 (2016).

\bibitem{Scherer2018}
Scherer, D.~D. \& Andersen, B.~M., Spin-Orbit Coupling and Magnetic Anisotropy
  in Iron-Based Superconductors. \textit{Phys. Rev. Lett.} \textbf{121}, 037205
  (2018).

\bibitem{Wang2017}
Wang, W., Park, J.~T., Yu, R., Li, Y., Song, Y., Zhang, Z., Ivanov, A., Kulda,
  J. \& Dai, P., Orbital selective neutron spin resonance in underdoped
  superconducting ${\mathrm{NaFe}}_{0.985}{\mathrm{Co}}_{0.015}\mathrm{As}$.
  \textit{Phys. Rev. B} \textbf{95}, 094519 (2017).

\bibitem{Zhang2016a}
Zhang, C., Lv, W., Tan, G., Song, Y., Carr, S.~V., Chi, S., Matsuda, M.,
  Christianson, A.~D., Fernandez-Baca, J.~A., Harriger, L.~W. \& Dai, P.,
  Electron doping evolution of the neutron spin resonance in
  ${\mathrm{NaFe}}_{1\ensuremath{-}x}{\mathrm{Co}}_{x}\mathrm{As}$.
  \textit{Phys. Rev. B} \textbf{93}, 174522 (2016).

\bibitem{Lee2016a}
Lee, C.~H., Kihou, K., Park, J.~T., Horigane, K., Fujita, K., Wa\ss{}er, F.,
  Qureshi, N., Sidis, Y., Akimitsu, J. \& Braden, M., Suppression of
  spin-exciton state in hole overdoped iron-based superconductors.
  \textit{Scientific Reports} \textbf{6}, 23424 (2016).

\bibitem{Horigane2016}
Horigane, K., Kihou, K., Fujita, K., Kajimoto, R., Ikeuchi, K., Ji, S.,
  Akimitsu, J. \& Lee, C.~H., Spin excitations in hole-overdoped iron-based
  superconductors. \textit{Scientific Reports} \textbf{6}, 33303 (2016).

\bibitem{Wang2013a}
Wang, M., Zhang, C.,  Lu, X., Tan, G., Luo, H., Song, Y., Wang, M., Zhang, X., Goremychkin, E.A., Perring, T.G., Maier, T.A., Yin, Z., Haule, K., Kotliar, G. \& Dai, P.,
Doping dependence of spin excitations and its correlations with high-temperature superconductivity in iron pnictides.
\textit{Nat Commun.} \textbf{4}, 2874 (2013).

\bibitem{Murai2018}
Murai, N., Suzuki, K., Ideta, S.-i., Nakajima, M., Tanaka, K., Ikeda, H. \&
  Kajimoto, R., Effect of electron correlations on spin excitation bandwidth in
  ${\mathrm{Ba}}_{0.75}{\mathrm{K}}_{0.25}{\mathrm{Fe}}_{2}{\mathrm{As}}_{2}$
  as seen via time-of-flight inelastic neutron scattering. \textit{Phys. Rev.
  B} \textbf{97}, 241112 (2018).

\bibitem{Guo2019}
Guo, J., Yue, L., Iida, K., Kamazawa, K., Chen, L., Han, T., Zhang, Y. \& Li,
  Y., Preferred Magnetic Excitations in the Iron-Based
  ${\mathrm{Sr}}_{1\ensuremath{-}x}{\mathrm{Na}}_{x}{\mathrm{Fe}}_{2}{\mathrm{As}}_{2}$
  Superconductor. \textit{Phys. Rev. Lett.} \textbf{122}, 017001 (2019).

\bibitem{Allred2015}
Allred, J.~M., Avci, S., Chung, D.~Y., Claus, H., Khalyavin, D.~D., Manuel, P.,
  Taddei, K.~M., Kanatzidis, M.~G., Rosenkranz, S., Osborn, R. \& Chmaissem,
  O., Tetragonal magnetic phase in
  ${\mathrm{Ba}}_{1\ensuremath{-}x}{\mathrm{K}}_{x}{\mathrm{Fe}}_{2}{\mathrm{As}}_{2}$
  from x-ray and neutron diffraction. \textit{Phys. Rev. B} \textbf{92}, 094515
  (2015).

\bibitem{Allred2016}
Allred, J.~M., Taddei, K.~M., Bugaris, D.~E., Krogstad, M.~J., Lapidus, S.~H.,
  Chung, D.~Y., Claus, H., Kanatzidis, M.~G., Brown, D.~E., Kang, J.,
  Fernandes, R.~M., Eremin, I., Rosenkranz, S., Chmaissem, O. \& Osborn, R.,
  Double-Q spin-density wave in iron arsenide superconductors. \textit{Nat.
  Phys.} \textbf{12}, 5, 493 (2016).

\bibitem{Mallett2015a}
Mallett, B. P.~P., Marsik, P., Yazdi-Rizi, M., Wolf, T., B\"ohmer, A.~E.,
  Hardy, F., Meingast, C., Munzar, D. \& Bernhard, C., Infrared Study of the
  Spin Reorientation Transition and Its Reversal in the Superconducting State
  in Underdoped
  ${\mathrm{Ba}}_{1-x}{\mathrm{K}}_{x}{\mathrm{Fe}}_{2}{\mathrm{As}}_{2}$.
  \textit{Phys. Rev. Lett.} \textbf{115}, 027003 (2015).

\bibitem{Wang2016}
Wang, L., Hardy, F., B\"ohmer, A.~E., Wolf, T., Schweiss, P. \& Meingast, C.,
  Complex phase diagram of
  ${\mathrm{Ba}}_{1\ensuremath{-}x}{\mathrm{Na}}_{x}{\mathrm{Fe}}_{2}{\mathrm{As}}_{2}$:
  A multitude of phases striving for the electronic entropy. \textit{Phys. Rev.
  B} \textbf{93}, 014514 (2016).

\bibitem{Frandsen2017}
Frandsen, B.~A., Taddei, K.~M., Yi, M., Frano, A., Guguchia, Z., Yu, R., Si,
  Q., Bugaris, D.~E., Stadel, R., Osborn, R., Rosenkranz, S., Chmaissem, O. \&
  Birgeneau, R.~J., Local Orthorhombicity in the Magnetic ${C}_{4}$ Phase of
  the Hole-Doped Iron-Arsenide Superconductor
  ${\mathrm{Sr}}_{1\ensuremath{-}x}{\mathrm{Na}}_{x}{\mathrm{Fe}}_{2}{\mathrm{As}}_{2}$.
  \textit{Phys. Rev. Lett.} \textbf{119}, 187001 (2017).

\bibitem{note-geom}
The geometry factors, $sin^2(\alpha)$ for detecting a $c$ polarized signal at
  (0.5,0.5,L) amount to 0.84, 0.37, 0.18, and 0.10 at L=1, 3, 5, and 7
  respectively. Those for an $a$ polarized signal are 1-$sin^2(\alpha)$.

\bibitem{suppl-inf}
Supplementary material containing further experimental details and INS data,
  ... .

\bibitem{Wasser2016}
Wa\ss{}er, F., Wurmehl, S., Aswartham, S., Sidis, Y., Park, J.~T.,
  Schneidewind, A., B\"{u}chner, B. \& Braden, M., Spin reorientation
  transition in Na-doped BaFe2As2 studied by single-crystal neutron
  diffraction. \textit{physica status solidi (b)} \textbf{254}, 1600181 (2017).

\bibitem{Pratt2010}
Pratt, D.~K., Kreyssig, A., Nandi, S., Ni, N., Thaler, A., Lumsden, M.~D.,
  Tian, W., Zarestky, J.~L., Bud'ko, S.~L., Canfield, P.~C., Goldman, A.~I. \&
  McQueeney, R.~J., Dispersion of the superconducting spin resonance in
  underdoped and antiferromagnetic \ce{BaFe2As2}. \textit{Phys. Rev. B}
  \textbf{81}, 140510 (2010).

\bibitem{steffens2011}
Steffens, P., Friedt, O., Sidis, Y., Link, P., Kulda, J., Schmalzl, K.,
  Nakatsuji, S. \& Braden, M., Magnetic excitations in the metallic
  single-layer ruthenates Ca$_{2-x}$Sr$_x$RuO$_4$ studied by inelastic neutron
  scattering. \textit{Phys. Rev. B} \textbf{83}, 054429 (2011).

\bibitem{Zanen}
Zaanen, J., Why the temperature is high. \textit{nature} \textbf{430}, 512
  (2004).


\bibitem{Scalapino}
Scalapino, D. J., A common thread: The pairing interaction for unconventional superconductors.
Rev. Mod. Phys. \textbf{84}, 1383 (2012).


\bibitem{note-exch}
{\red The exchange energy can be estimated within the Heisenberg model, $H=\sum_{ij}{J_{ij}<S_iS_j>}$ \cite{Scalapino} by
using the $J_{ij}S$ and $S$ values of pure BaFe$_2$As$_2$ \cite{Qureshi2012,Wang2013}.
However a better characterization of the sample, temperature and energy dependent $Q$ widths is required for a
precise determination. With the simplified suceptibility model of constant $Q$ widths and assuming
fully isotropic fluctuations (except SRM-1) one finds a gain in exchange energy for Na40 about an order of magnitude larger than
the condensation energy determined from specific heat measurements on a single crystal of Ba$_{1.65}$Na$_{0.35}$Fe$_2$As$_2$,
0.05\,meV/Fe \cite{Pramanik2011}. This large ratio between exchange energy gain and condensation energy agrees with
reports for Ba$_{1.67}$K$_{0.33}$Fe$_2$As$_2$ \cite{Wang2013a} as well as for cuprates and  CeCu$_2$Si$_2$ \cite{Scalapino}.
}

\bibitem{Pramanik2011}
Pramanik, A.~K., Abdel-Hafiez, M., Aswartham, S., Wolter, A. U.~B., Wurmehl,
  S., Kataev, V. \& B\"uchner, B., Multigap superconductivity in single
  crystals of \ce{Ba_{0.65}Na_{0.35}Fe_2As_2}: A calorimetric investigation.
  \textit{Phys. Rev. B} \textbf{84}, 064525 (2011).

\bibitem{Ismer2007}
Ismer, J.-P., Eremin, I., Rossi, E. \& Morr, D.~K., Magnetic Resonance in the
  Spin Excitation Spectrum of Electron-Doped Cuprate Superconductors.
  \textit{Phys. Rev. Lett.} \textbf{99}, 047005 (2007).

\bibitem{Rowe2012}
Rowe, W., Knolle, J., Eremin, I. \& Hirschfeld, P.~J., Spin excitations in
  layered antiferromagnetic metals and superconductors. \textit{Phys. Rev. B}
  \textbf{86}, 134513 (2012).

\bibitem{Aswartham2012}
Aswartham, S., Abdel-Hafiez, M., Bombor, D., Kumar, M., Wolter, A. U.~B., Hess,
  C., Evtushinsky, D.~V., Zabolotnyy, V.~B., Kordyuk, A.~A., Kim, T.~K.,
  Borisenko, S.~V., Behr, G., B\"uchner, B. \& Wurmehl, S., Hole doping in
  BaFe$_2$As$_2$: The case of Ba$_{1-x}$Na$_x$Fe$_2$As$_2$ single crystals.
  \textit{Phys. Rev. B} \textbf{85}, 224520 (2012).

\bibitem{Qureshi2014a}
Qureshi, N., Steffens, P., Lamago, D., Sidis, Y., Sobolev, O., Ewings, R.~A.,
  Harnagea, L., Wurmehl, S., B\"uchner, B. \& Braden, M., Fine structure of the
  incommensurate antiferromagnetic fluctuations in single-crystalline LiFeAs
  studied by inelastic neutron scattering. \textit{Phys. Rev. B} \textbf{90},
  144503 (2014).

\bibitem{Zheludev2006}
Zheludev, A., \textit{Manual: ResLib 3.4}. Neutron Scattering Sciences
  Division, Oak Ridge National Laboratory, Oak Ridge, TN 37831-6393, USA
  (2007).

\bibitem{Tucker2014a}
Tucker, G.~S., Fernandes, R.~M., Pratt, D.~K., Thaler, A., Ni, N., Marty, K.,
  Christianson, A.~D., Lumsden, M.~D., Sales, B.~C., Sefat, A.~S., Bud'ko,
  S.~L., Canfield, P.~C., Kreyssig, A., Goldman, A.~I. \& McQueeney, R.~J.,
  Crossover from spin waves to diffusive spin excitations in underdoped
  Ba(Fe$_{1-x}$Co$_x$)$_2$As$_2$. \textit{Phys. Rev. B} \textbf{89}, 180503
  (2014).

\bibitem{Xu2013}
Xu, G., Xu, Z. \& Tranquada, J.~M., Absolute cross-section normalization of
  magnetic neutron scattering data. \textit{Rev. Sci. Instrum.} \textbf{84},
  083906 (2013).

\end{thebibliography}
\end{document}